\documentclass[pdflatex,sn-mathphys-num]{sn-jnl}

\usepackage{graphicx}
\usepackage{amsmath,amssymb,amsfonts}
\usepackage{bm}

\newcommand{\ii}{\mathrm{i}}

\begin{document}

\title[Adiabatic particle production in a time-dependent accelerated frame]{Adiabatic particle production in a time-dependent accelerated frame}

\author*[1]{\fnm{V\'ictor M.} \sur{Villalba}}\email{victor.villalba2@monash.edu}

\affil*[1]{\orgdiv{School of Mathematics}, \orgname{Monash University}, \orgaddress{\city{Clayton}, \postcode{3800}, \state{VIC}, \country{Australia}}}

\abstract{We study a massive scalar field in flat spacetime written in a time-dependent accelerated frame whose fixed-coordinate observers have maximal proper acceleration at $t=0$ and become asymptotically inertial without forming a horizon. The Klein--Gordon equation separates into a modified Mathieu equation in time and a confining Schr\"odinger problem in space with a discrete spectrum. We take as initial state the adiabatic vacuum of the frame at the instant of maximal acceleration and compute the particle number in the inertial future. Analytic continuation of the spatial eigenfunctions gives the out modes in closed form, and we show that they are of positive frequency with respect to Minkowski time, so that the out vacuum is the Minkowski vacuum. The particle number therefore counts the inertial quanta contained in the frame's instantaneous vacuum, and is given exactly by the mismatch at $t=0$ between the frequency of the initial state and an eigenvalue of the Minkowski one-particle operator. For the zeroth-order state it falls as the inverse eighth power of the initial mode frequency and grows as the fourth power of the maximal acceleration; raising the adiabatic order cancels successive terms of the endpoint expansion, in analogy with the switching transients of finite-time detectors. For this state the occupation numbers decrease algebraically, not exponentially, with frequency.}


\maketitle

\section{Introduction}
\label{sec:intro}

Quantum field theory does not assign a unique particle content to a given state: the notion of particle depends on the choice of time and of positive-frequency modes, and hence on the observer \cite{Fulling1973,Davies1975,Unruh1976,BirrellDavies1982}. The best-known example is the Unruh effect. A uniformly accelerated observer, whose natural coordinates are those of the Rindler frame \cite{Rindler1966}, perceives the Minkowski vacuum as a thermal bath at temperature $\mathfrak a/2\pi$ \cite{Unruh1976,UnruhWald1984}, where $\mathfrak a$ is its proper acceleration; see Refs.~\cite{Takagi1986,Crispino2008} for reviews. Uniform acceleration is, however, an idealisation. It lasts forever and is accompanied by a Rindler horizon, whereas physically realisable observers accelerate for a finite time and with an acceleration that changes.

Departures from uniform acceleration have been studied mostly through the response of Unruh--DeWitt detectors. Obadia and Milgrom analysed general trajectories \cite{ObadiaMilgrom2007}, while Kothawala and Padmanabhan \cite{KothawalaPadmanabhan2010} and Barbado and Visser \cite{BarbadoVisser2012} considered detectors whose acceleration depends on time; the entanglement of detectors on non-uniformly accelerated trajectories was studied in Ref.~\cite{Ostapchuk2012}. Thermal-like responses without a Rindler horizon have also been found, for a uniformly accelerated detector whose horizon is hidden by a mirror \cite{Nicolaevici2015} and for detectors with time-dependent acceleration \cite{Sobrero2025}. Conversely, Mu\~noz and Doria \cite{MunozDoria2026} showed that for asymptotically inertial observers with non-constant acceleration and no horizon, the Minkowski vacuum is a pure single-mode squeezed state of the observers' modes: Bose--Einstein-like occupation numbers can occur without a thermal density matrix. A recurring lesson of this literature is that finite-time effects require care: the response of a detector switched on for a finite time contains transient contributions associated with the switching, which can dominate unless the switching is smooth \cite{SriramkumarPadmanabhan1996,LoukoSatz2006,Satz2007}. The onset of the Hawking--Unruh response \cite{JuarezAubryLouko2014} and the time a detector must interact with the field before its response appears thermal \cite{Fewster2016,Parry2025} have been analysed in detail, and the comparison between inertial and accelerated descriptions of a finite-time measurement has recently been revisited \cite{Dickinson2025}; the transient terms in the finite-time Unruh response, and the time after which they become negligible, were analysed in Ref.~\cite{Stargen2026}. In flat spacetime, particle creation driven by time dependence also arises for moving mirrors \cite{FullingDavies1976,DaviesFulling1977}, whose time-dependent particle production was followed in Ref.~\cite{GoodAndersonEvans2013}, and for time-dependent potentials that mimic gravitational backgrounds \cite{DuarteBaptista2026}.

A complementary approach works directly with the field modes in accelerated coordinates and computes the Bogoliubov coefficients between the natural mode bases \cite{BirrellDavies1982}. Separable coordinate systems adapted to accelerated observers were constructed by Costa \cite{Costa1989}, and particle creation in frames that become uniformly accelerated at late times was studied in Refs.~\cite{PercocoVillalba1992,VillalbaMateu2000}. In those frames the late-time observers approach the setting of the Unruh effect. Here we consider the opposite situation: a frame whose acceleration is largest at a finite time and decays to zero, so that the observers end in inertial motion and no horizon forms. Related Bogoliubov calculations have been carried out between successive Rindler frames \cite{Dubey2026} and between the inertial and accelerated phases of a rigid cavity \cite{ShahKolekar2026}; in the latter case the excitation spectrum is a power law set by the acceleration. We use the exactly flat metric (units $\hbar=c=1$, signature $+---$)
\begin{align}
ds^2&=F(t,x)\,(dt^2-dx^2)-dy^2-dz^2,\nonumber\\
F&=C\,[\cosh\kappa x+\cosh\kappa t],
\label{eq:metric}
\end{align}
with $C,\kappa>0$. The coordinate transformation given in Sect.~\ref{sec:geometry} shows that this metric is Minkowski spacetime written in accelerated coordinates. For $\kappa=2a$ and $C=1/2$, with $a$ the acceleration parameter of Ref.~\cite{Sobrero2025}, it coincides with the chart used by Sobrero et al., who computed the response of an Unruh--DeWitt detector at fixed $x$ in the Minkowski vacuum. The present work is complementary: we work with the field modes and Bogoliubov coefficients of the frame rather than with a detector. Fixed-$x$ observers have their maximum proper acceleration at $t=0$; for $t>0$ the acceleration decreases monotonically, falling as the inverse cube of proper time, and the observers approach inertial motion with a finite terminal velocity. We consider the half-line evolution $0\le t<\infty$, with the state prescribed at $t=0$ and the particle number read in the asymptotically inertial future. Since $F$ is even in $t$, the observers are also asymptotically inertial in the past; we do not consider that in--out problem here.

Because the evolution starts at a finite time, the initial state must be specified there. The natural choice is an adiabatic vacuum \cite{Parker1969,ParkerFulling1974,BirrellDavies1982,ParkerToms2009,LudersRoberts1990}, whose lowest order is the instantaneous vacuum \cite{Agullo2015,Bianchi2026}. Its definition depends on both the time at which it is imposed and its adiabatic order \cite{Fulling1979,Parker2012}, and the time dependence and physical meaning of the adiabatic particle number have been clarified in the Schwinger and de Sitter problems \cite{Dabrowski2014,Dabrowski2016,Ilderton2022}. Prescribing the state at $t=0$ is the field-mode counterpart of switching on a detector at a finite time. We show that, as for detectors, the leading particle number is a transient set by the initial instant, and we determine how it depends on the adiabatic order of the initial state.

Our main results are as follows. The Klein--Gordon equation separates into a modified Mathieu equation in $t$ and a confining Schr\"odinger problem in $x$. The physical separation constants are discrete and satisfy $\Lambda_n<-A$, so every physical temporal mode has positive frequency on the real half-line. For the zeroth-order adiabatic vacuum imposed at $t=0$, the future particle number has an endpoint expansion in inverse powers of $\gamma=\sqrt A/\kappa$, with leading term
\begin{equation}
N_{\rm out}\simeq\frac{\kappa^4A^2}{256\,\Omega_0^8}.
\end{equation}
Direct numerical integration confirms this law for the physical modes. Increasing the adiabatic order removes successive endpoint terms, producing stronger algebraic suppression in the adiabatic large-$\gamma$ limit. For this state the occupation numbers decrease algebraically with frequency, not exponentially.

We then show that the spatial eigenfunctions, continued to ${\rm Im}\,x=\pi/\kappa$, are exactly the positive-frequency out modes. This gives the particle number in closed form,
\begin{equation}
N_{\rm out}=\frac{(W_0-p_n)^2}{4W_0\,p_n},
\end{equation}
where $W_0$ is the frequency of the initial adiabatic state and $p_n$ is a number obtained from an ordinary Mathieu equation. The endpoint series is the expansion of this mismatch in $1/\gamma$. We prove that the out modes are of positive frequency with respect to Minkowski time, so that the out vacuum is the Minkowski vacuum and $p_n$ is an eigenvalue of the Minkowski one-particle operator on the slice $t=0$. The particle number is thus the number of inertial quanta in the instantaneous vacuum of the accelerated frame, and the acceleration enters through the position dependence of the lapse of the frame at $t=0$.

The paper is organised as follows. Section~\ref{sec:geometry} describes the geometry of the accelerated frame, and Sect.~\ref{sec:KG} separates the field equation and determines the mode spectrum. Section~\ref{sec:vacuum} defines the adiabatic initial state and the Bogoliubov coefficients, and Sect.~\ref{sec:endpoint} derives the endpoint expansion, its dependence on the adiabatic order, and the exact result for the particle number. Numerical results are presented in Sect.~\ref{sec:numerics}, and Sects.~\ref{sec:discussion} and \ref{sec:conclusions} contain the discussion and conclusions.

\section{Time-dependent accelerated coordinates}
\label{sec:geometry}

We first construct the transformation to Minkowski coordinates and then describe the worldlines of the fixed-$x$ observers. In null coordinates $u=t-x$, $v=t+x$,
\begin{equation}
\cosh\kappa t+\cosh\kappa x
=2\cosh\frac{\kappa u}{2}\cosh\frac{\kappa v}{2},
\end{equation}
so that $F=2C\cosh(\kappa u/2)\cosh(\kappa v/2)$. Define
\begin{equation}
\bar u=\ell\sinh\frac{\kappa u}{2},\qquad
\bar v=\ell\sinh\frac{\kappa v}{2},\qquad \ell\equiv\frac{2\sqrt{2C}}{\kappa}.
\end{equation}
Then $d\bar u\,d\bar v=F\,du\,dv$. With $\bar t=(\bar u+\bar v)/2$ and $\bar x=(\bar v-\bar u)/2$,
\begin{equation}
\bar t=\ell\sinh\frac{\kappa t}{2}\cosh\frac{\kappa x}{2},\qquad
\bar x=\ell\cosh\frac{\kappa t}{2}\sinh\frac{\kappa x}{2},
\label{eq:embedding}
\end{equation}
and therefore
\begin{equation}
d\bar t^2-d\bar x^2=F(t,x)(dt^2-dx^2).
\end{equation}
Thus Eq.~\eqref{eq:metric} is the Minkowski metric written in accelerated coordinates. Since $\sinh$ is one-to-one on $\mathbb R$, the transformation is global in the $(t,x)$ plane. The slice $t=0$ is the inertial slice $\bar t=0$, on which $\bar x=\ell\sinh(\kappa x/2)$: the initial surface of the present problem is a surface of simultaneity of an inertial frame.

\begin{figure}[t]
\includegraphics[width=0.95\textwidth]{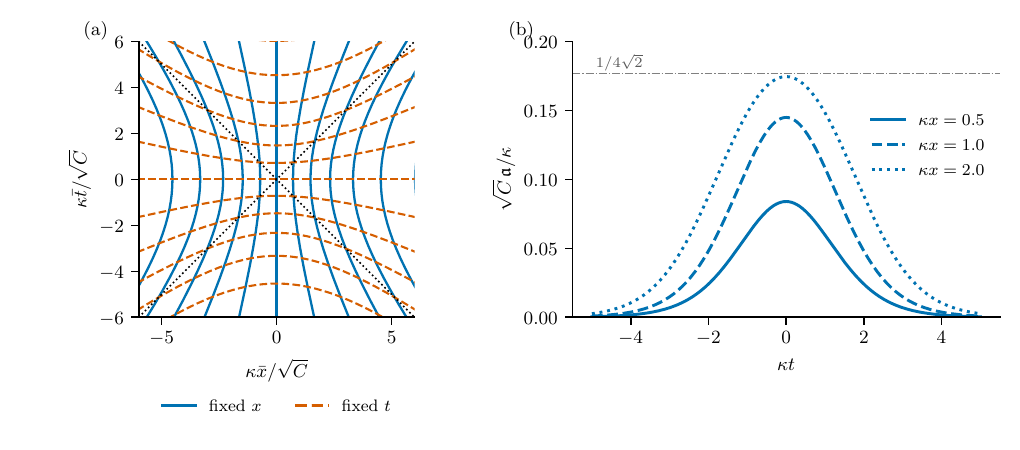}
\caption{(a) Worldlines of fixed $x$ (solid) and surfaces of constant $t$ (dashed) in the inertial coordinates \eqref{eq:embedding}, both families of hyperbolae \eqref{eq:hyperbolae}; the dotted lines are the light cone $\bar t=\pm\bar x$. (b) Proper acceleration \eqref{eq:accel} of fixed-$x$ observers. The acceleration is maximal at $t=0$ and tends to zero for $t\to+\infty$.}
\label{fig:geometry}
\end{figure}

Eliminating $t$ or $x$ from Eq.~\eqref{eq:embedding} gives
\begin{equation}
\frac{\bar x^2}{\ell^2\sinh^2(\kappa x/2)}-\frac{\bar t^2}{\ell^2\cosh^2(\kappa x/2)}=1,\qquad
\frac{\bar t^2}{\ell^2\sinh^2(\kappa t/2)}-\frac{\bar x^2}{\ell^2\cosh^2(\kappa t/2)}=1 .
\label{eq:hyperbolae}
\end{equation}
The worldline of an observer at fixed $(x,y,z)$, $x\neq0$, is therefore a timelike hyperbola with vertex at
$\bar x=\ell\sinh(\kappa x/2)$ on $\bar t=0$ and asymptotes $\bar x=\pm\tanh(\kappa x/2)\,\bar t$, and the
slices of constant $t\neq0$ are spacelike hyperbolae (Fig.~\ref{fig:geometry}(a)). The observer at $x=0$ is
inertial and sits at $\bar x=0$. In contrast with the Rindler hyperbolae $\bar x^2-\bar t^2={\rm const}$, the
asymptotes lie strictly inside the light cone.

Along a fixed-$x$ worldline the velocity and rapidity in the inertial frame are
\begin{equation}
v=\frac{d\bar x}{d\bar t}=\tanh\frac{\kappa t}{2}\,\tanh\frac{\kappa x}{2},\qquad \eta={\rm artanh}\,v,
\label{eq:vel}
\end{equation}
and the proper time is $d\tau_{\rm p}=\sqrt F\,dt$. Every observer is at rest at $t=0$ and approaches the
terminal velocity $v_\infty=\tanh(\kappa x/2)$ as $t\to+\infty$. For rectilinear motion the magnitude of the proper acceleration
is $\mathfrak a=|d\eta/d\tau_{\rm p}|$. Using
$\cosh^2\frac{\kappa t}{2}\cosh^2\frac{\kappa x}{2}-\sinh^2\frac{\kappa t}{2}\sinh^2\frac{\kappa x}{2}=F/2C$,
\begin{equation}
\mathfrak a=\frac{1}{\sqrt F}\Big|\frac{d\eta}{dt}\Big|
=\frac{\kappa C|\sinh\kappa x|}{2F^{3/2}}
=\frac{\kappa|\sinh\kappa x|}{2\sqrt{C}\,[\cosh\kappa x+\cosh\kappa t]^{3/2}},
\label{eq:accel}
\end{equation}
which coincides with $|\partial_xF|/2F^{3/2}$ obtained from the four-acceleration of
$u^\mu=(F^{-1/2},0,0,0)$. Three properties of Eq.~\eqref{eq:accel} are used below.

(i) For fixed $x\neq0$, $\mathfrak a$ is maximal at $t=0$ and decreases monotonically for $t>0$. Its initial value
is $\mathfrak a_0(x)=\kappa\sinh(\kappa|x|/2)/[2\sqrt{2C}\cosh^2(\kappa x/2)]$, and over all observers
$\mathfrak a\le\kappa/(4\sqrt{2C})$, with equality at $\sinh(\kappa|x|/2)=1$, $t=0$.

(ii) Since $\mathfrak a=|d\eta/d\tau_{\rm p}|$ and $d\eta/d\tau_{\rm p}$ has the fixed sign of $x$, the acceleration accumulated over the future evolution is finite,
\begin{equation}
\int_0^\infty\mathfrak a\,d\tau_{\rm p}=|\eta(+\infty)-\eta(0)|=\frac{\kappa|x|}{2},
\end{equation}
so each observer acquires a rapidity of finite magnitude $\kappa|x|/2$ and ends with velocity $v_\infty$.

(iii) As $t\to+\infty$, $\sqrt F\simeq\sqrt{C/2}\,e^{\kappa t/2}$ and $\tau_{\rm p}\simeq\sqrt{2C}\,e^{\kappa t/2}/\kappa$, so
\begin{equation}
\mathfrak a\simeq\frac{\sqrt2\,\kappa|\sinh\kappa x|}{\sqrt C}\,e^{-3\kappa t/2}
\simeq\frac{4C|\sinh\kappa x|}{\kappa^2\tau_{\rm p}^3}.
\label{eq:latea}
\end{equation}
The acceleration decays as the inverse cube of proper time, and the observers become asymptotically inertial.
Since the asymptotic velocity is subluminal, the causal past of every fixed-$x$ worldline is the whole
spacetime, and no Rindler horizon forms.

At late times the frame approaches a Milne frame. From Eq.~\eqref{eq:embedding},
$\bar t^2-\bar x^2=\ell^2[\sinh^2(\kappa t/2)-\sinh^2(\kappa x/2)]$, and for $t\to+\infty$
\begin{equation}
\bar t\simeq\frac{\ell}{2}e^{\kappa t/2}\cosh\frac{\kappa x}{2},\qquad
\bar x\simeq\frac{\ell}{2}e^{\kappa t/2}\sinh\frac{\kappa x}{2}.
\label{eq:milne}
\end{equation}
The fixed-$x$ observers therefore end as the comoving observers of the Milne wedge, the future light cone of the origin, with Milne time $T=\sqrt{\bar t^2-\bar x^2}\simeq(\ell/2)\,e^{\kappa t/2}=\sqrt{2C}\,e^{\kappa t/2}/\kappa$ and rapidity $\chi=\kappa x/2$. Correspondingly, $F(dt^2-dx^2)\to dT^2-T^2d\chi^2$, and $T$ coincides asymptotically with the proper time of item (iii).

The present coordinates belong to the family of flat accelerated systems that includes Rindler and Costa-type coordinates \cite{Costa1989}. What distinguishes them is that the proper acceleration is time dependent, with a finite maximum at $t=0$ and finite total rapidity.

\section{Field equation and mode structure}
\label{sec:KG}

\subsection{Separation and spectrum}

The inverse metric is $g^{\mu\nu}={\rm diag}(F^{-1},-F^{-1},-1,-1)$ and $\sqrt{-g}=F$, so
$(\Box+m^2)\Phi=0$ becomes
\begin{equation}
\big[\partial_t^2-\partial_x^2-F(\partial_y^2+\partial_z^2)+m^2F\big]\Phi=0.
\end{equation}
With $\Phi=e^{\ii(k_yy+k_zz)}\psi(t,x)$ and $\mu^2=m^2+k_y^2+k_z^2$, this reads
$[\partial_t^2-\partial_x^2+\mu^2F]\psi=0$. For $\mu=0$ (a massless field with $k_y=k_z=0$) the field is
conformally trivial in the $(t,x)$ plane and no particles are produced, so we take $\mu>0$. Set
$A\equiv\mu^2C$ and $\psi=T(t)X(x)$. Since
$F$ is a sum of a function of $t$ and a function of $x$, the equation separates with constant $\Lambda$:
\begin{align}
\ddot T+\big[A\cosh\kappa t-\Lambda\big]T&=0, \label{eq:T}\\
X''-\big[A\cosh\kappa x+\Lambda\big]X&=0 . \label{eq:X}
\end{align}
With $x_M=\kappa t/2$, Eq.~\eqref{eq:T} becomes
$T_{x_Mx_M}+[2q_M\cosh2x_M-a_M]T=0$ with
\begin{equation}
q_M=\frac{2A}{\kappa^2}=2\gamma^2,\quad a_M=\frac{4\Lambda}{\kappa^2}=4\gamma^2 r,
\end{equation}
where $\gamma\equiv\sqrt A/\kappa$ and $r\equiv\Lambda/A$. This is the modified Mathieu equation \cite[Sec.~28.20]{NISTHandbook}. Mathieu-type mode equations also arise in parametric resonance after inflation \cite{KLS1997}, where the frequency is periodic in time and resonance bands produce exponential particle growth. Equation~\eqref{eq:T} has no such bands: for $t>0$ the squared frequency increases monotonically. The Klein--Gordon inner product on a
$t=$const slice, $(\psi_1,\psi_2)=\ii\int dx\,d^2y\,(\psi_1^*\partial_t\psi_2-\partial_t\psi_1^*\psi_2)$,
carries no conformal factor because $\sqrt{-g}\,g^{tt}=1$.

\label{sec:spectrum}

Equation \eqref{eq:X} is a Schr\"odinger problem $H_xX=EX$ with
$H_x=-\partial_x^2+A\cosh\kappa x$ and $E=-\Lambda$. Because the norm above is the $L^2(dx)$ norm
of the mode data, physical modes have $X\in L^2(\mathbb R)$. Multiplying \eqref{eq:X} by
$X^*$ and integrating,
\begin{align}
-\Lambda\!\int\!|X|^2dx&=\int\!\big(|X'|^2+A\cosh\kappa x\,|X|^2\big)dx\nonumber\\
&>A\!\int\!|X|^2dx ,
\end{align}
so that $E>A$, i.e.\
\begin{equation}
\Lambda_n<-A\quad\Longleftrightarrow\quad r_n<-1.
\end{equation}
Since $A\cosh\kappa x\to\infty$ as $|x|\to\infty$, the spectrum is discrete. We write $\psi=\sum_nT_n(t)X_n(x)$, with $H_xX_n=E_nX_n$.

\subsection{Frequency, adiabaticity and future modes}

Equation \eqref{eq:T} is $\ddot T+\Omega^2T=0$ with $\Omega^2(t)=A\cosh\kappa t-\Lambda
=A(\cosh\kappa t-r)$. For all physical modes $\Omega^2\ge A+E_n>0$, so there is no real turning
point, and $\Omega_0\equiv\Omega(0)=\sqrt{A-\Lambda}=\sqrt{A(1-r)}$. Since $\Omega$ is even,
$\dot\Omega(0)=0$. The adiabaticity parameter is
\begin{equation}
\epsilon(t)=\Big|\frac{\dot\Omega}{\Omega^2}\Big|=\frac{A\kappa|\sinh\kappa t|}{2[A\cosh\kappa t-\Lambda]^{3/2}}
\;\xrightarrow{t\to\infty}\;\frac{\kappa\,e^{-\kappa t/2}}{\sqrt{2A}},
\end{equation}
so the evolution becomes increasingly adiabatic at late times. Since $\epsilon(0)=0$ and $\epsilon$ decays as $e^{-\kappa t/2}$, the non-adiabatic stage is the interval $\kappa t=O(1)$ that follows the initial time.

For $t\to\infty$, $\Omega^2=\frac A2e^{\kappa t}-\Lambda+\frac A2e^{-\kappa t}$, and the last term
is exponentially small. With $\xi=\sqrt2\,\gamma\,e^{\kappa t/2}$ and
$d^2/dt^2=(\kappa^2/4)(z^2\partial_z^2+z\partial_z)$ for $z=e^{\kappa t/2}$, Eq.~\eqref{eq:T}
becomes Bessel's equation
\begin{equation}
\xi^2T_{\xi\xi}+\xi T_\xi+(\xi^2-\nu^2)T=0,\qquad \nu^2=\frac{4\Lambda}{\kappa^2},\quad \nu=2\gamma\sqrt r .
\end{equation}
For the physical modes $\Lambda<0$, so $\nu=\ii\sigma_n$ with $\sigma_n=2\sqrt{E_n}/\kappa$, as for
Rindler modes. The corresponding asymptotic positive- and negative-frequency modes are
\begin{align}
u_+&=\mathcal N_2H^{(2)}_\nu(\xi),\quad u_-=\mathcal N_1H^{(1)}_\nu(\xi),\nonumber\\
\mathcal N_{2,1}&=\sqrt{\tfrac{\pi}{2\kappa}}\,e^{\mp\ii(\nu\pi/2+\pi/4)},
\label{eq:out}
\end{align}
normalised so that $u_\pm\to(2\Omega)^{-1/2}e^{\mp\ii\Theta}$ up to a constant phase, as follows
from the large-argument form $H^{(2,1)}_\nu(\xi)\simeq\sqrt{2/\pi\xi}\,e^{\mp\ii(\xi-\nu\pi/2-\pi/4)}$
and $\Omega\to\kappa\xi/2$. With $W\{H^{(1)}_\nu,H^{(2)}_\nu\}=-4\ii/\pi\xi$ \cite[Sec.~10.5]{NISTHandbook}
one has $u_+\dot u_- - \dot u_+u_-=\ii$ for real and imaginary $\nu$.

\section{Adiabatic vacuum and particle number}
\label{sec:vacuum}

The state is specified at $t=0$ and evolved only forward over the half-line $0\le t<\infty$. The particle number is defined with respect to the adiabatic out basis in the future, where the proper acceleration tends to zero and the evolution becomes asymptotically adiabatic.

\subsection{Adiabatic expansion and initial state}
\label{sec:adexp}

Each temporal mode obeys $\ddot T+\Omega^2(t)T=0$ and can be viewed as a harmonic oscillator with
Hamiltonian $H(t)=\tfrac12\big(p^2+\Omega^2(t)q^2\big)$. An exact solution may always be written in
the WKB form \cite{ParkerFulling1974,BirrellDavies1982}
\begin{equation}
T(t)=\frac{1}{\sqrt{2W(t)}}\,\exp\!\Big(-\ii\!\int^t W(s)\,ds\Big),
\label{eq:wkb}
\end{equation}
which satisfies the Wronskian condition $T\dot T^*-T^*\dot T=\ii$ identically, provided $W$ obeys
\begin{equation}
W^2=\Omega^2-\frac12\Big[\frac{\ddot W}{W}-\frac32\Big(\frac{\dot W}{W}\Big)^2\Big].
\label{eq:Weq}
\end{equation}
The adiabatic expansion solves \eqref{eq:Weq} iteratively, assigning adiabatic order $n$ to a term
with $n$ time derivatives. Formally one replaces $\Omega(t)\to\Omega(\varepsilon t)$ and expands in
$\varepsilon$; only even orders appear, and
\begin{equation}
W_{(0)}=\Omega,\qquad
W^2_{(2)}=\Omega^2-\frac12\frac{\ddot\Omega}{\Omega}+\frac34\Big(\frac{\dot\Omega}{\Omega}\Big)^2 ,
\label{eq:W2}
\end{equation}
and so on. The series is asymptotic, not convergent. For the cosh profile, each time derivative of
$\Omega=\sqrt A\,(\cosh\kappa t-r)^{1/2}$ brings a factor $\kappa$, so the expansion parameter is
$\kappa/\sqrt A=1/\gamma$, and the adiabatic expansion is an expansion in $1/\gamma$. This is the same
parameter that controls the weak-mixing integral of Sect.~\ref{sec:weak} and the endpoint series of
Sect.~\ref{sec:endpoint}.

The $n$th-order adiabatic vacuum at time $t_0$ is defined by requiring that the mode function and its
derivative coincide at $t_0$ with the truncated WKB form built from $W_{(n)}$
\cite{Parker1969,ParkerFulling1974,LudersRoberts1990}. Two different choices, $(n,t_0)$ and
$(n',t_0')$, generally define different states, related by a Bogoliubov transformation. In the present problem the state is imposed at $t=0$, the instant at which the frame is maximally accelerated, where $\dot\Omega(0)=0$ but $\ddot\Omega(0)=A\kappa^2/(2\Omega_0)\neq0$. The future is asymptotically adiabatic, so the corresponding out particle number is well defined.

The zeroth-order vacuum has a simple physical interpretation. The instantaneous annihilation
operator $a(t)=\big[\Omega(t)q+\ii p\big]/\sqrt{2\Omega(t)}$ diagonalises $H(t)$, and the
zeroth-order adiabatic vacuum at $t$ is the ground state of $H(t)$. The associated particle number
$N(t)=\langle a^\dagger(t)a(t)\rangle$ counts instantaneous oscillator quanta in the state evolved
from the initial state specified at $t=0$. This is the particle notion discussed by Fulling \cite{Fulling1979}: it is
well defined at every $t$ but depends on the choice of canonical variables used to write $H(t)$, and
only its $t\to\infty$ limit has an unambiguous meaning when the future is adiabatic
\cite{Parker2012,Dabrowski2016}.

We take as initial state the zeroth-order adiabatic vacuum at $t=0$,
\begin{equation}
T(0)=\frac1{\sqrt{2\Omega_0}},\qquad \dot T(0)=-\ii\Omega_0\,T(0),
\label{eq:ic}
\end{equation}
that is, Eq.~\eqref{eq:wkb} with $W=W_{(0)}=\Omega$ at $t_0=0$. It satisfies the Wronskian
normalisation $T\dot T^*-T^*\dot T=\ii$. The derivative of the zeroth-order WKB mode also contains
the term $\dot\Omega/(2\Omega)$ through $\dot T=-(\ii\Omega+\dot\Omega/2\Omega)T$; since
$\dot\Omega(0)=0$, this contribution vanishes at $t=0$. The first correction to the initial state therefore
occurs at second adiabatic order, Eq.~\eqref{eq:W2}, discussed in Sect.~\ref{sec:order}.

\subsection{Bogoliubov coefficients}

Write $T=(2\Omega)^{-1/2}[\alpha e^{-\ii\Theta}+\beta e^{\ii\Theta}]$ with $\Theta=\int_0^t\Omega\,ds$ and
the standard condition $\dot T=-\ii(\Omega/2)^{1/2}[\alpha e^{-\ii\Theta}-\beta e^{\ii\Theta}]$, which
expands the exact solution at each instant in the zeroth-order adiabatic basis. Differentiating the
first relation and comparing with the second gives
$\dot\alpha e^{-\ii\Theta}+\dot\beta e^{\ii\Theta}=(\dot\Omega/2\Omega)(\alpha e^{-\ii\Theta}+\beta e^{\ii\Theta})$,
while differentiating the second and using $\ddot T=-\Omega^2T$ gives the same relation with the
signs of the $\beta$ terms reversed and the right-hand side negated. Adding and subtracting,
\begin{equation}
\dot\alpha=\frac{\dot\Omega}{2\Omega}e^{2\ii\Theta}\beta,\qquad
\dot\beta=\frac{\dot\Omega}{2\Omega}e^{-2\ii\Theta}\alpha,
\label{eq:bog}
\end{equation}
with $\alpha(0)=1$, $\beta(0)=0$ for the state \eqref{eq:ic}, and $|\alpha|^2-|\beta|^2=1$, which is
conserved because the two cross terms in $\frac{d}{dt}(|\alpha|^2-|\beta|^2)$ cancel. The
zeroth-order adiabatic number is $N(t)=|\beta(t)|^2$. (To avoid confusion with the proper
acceleration $\mathfrak a$ of Sect.~\ref{sec:geometry}, $\alpha$ and $\beta$ denote Bogoliubov coefficients
throughout.) The coupling $\dot\Omega/2\Omega$ in
\eqref{eq:bog} is half the adiabaticity parameter times $\Omega$, so mixing between positive and
negative frequency occurs only where $\epsilon(t)$ is appreciable, i.e.\ for $\kappa t=O(1)$.

The instantaneous adiabatic number also has a useful finite-time interpretation. If the frequency were frozen at a time $t_1$, the instantaneous coefficients $\alpha(t_1)$ and $\beta(t_1)$ would become the corresponding final Bogoliubov coefficients. Thus $N(t_1)=|\beta(t_1)|^2$ may be viewed as the final yield of this auxiliary frozen-frequency problem, in analogy with the interpretation discussed by Ilderton \cite{Ilderton2022}. In the present problem $N(t)$ measures the particle content obtained by evolving the state specified at $t=0$ up to time $t$ and comparing it with the instantaneous adiabatic basis. Since the future is asymptotically adiabatic, $N(t)$ approaches the unambiguous final value $N_{\rm out}$ as $t\to+\infty$. Changing the adiabatic order changes the finite-time initial state, and therefore changes the algebraic endpoint contribution.

\label{sec:weak}

If $|\beta|\ll1$ then $|\alpha|^2=1+|\beta|^2\simeq1$, and Eq.~\eqref{eq:bog} gives
\begin{equation}
\beta_{\rm out}\simeq\int_0^\infty\frac{\dot\Omega}{2\Omega}e^{-2\ii\Theta}dt .
\end{equation}
With $\tau=\kappa t$, $\Omega=\sqrt A\sqrt{\cosh\tau-r}$ and $\Theta=\gamma S(\tau)$,
$S(\tau)=\int_0^\tau\sqrt{\cosh u-r}\,du$, this becomes
\begin{equation}
\beta_{\rm out}\simeq\int_0^\infty\phi(\tau)\,e^{-2\ii\gamma S(\tau)}d\tau,\qquad
\phi(\tau)=\frac{\sinh\tau}{4(\cosh\tau-r)}.
\label{eq:master}
\end{equation}
The parameter $\gamma$ multiplies a purely geometric phase, and large $\gamma$ is the
adiabatic limit.

\section{Endpoint asymptotics}
\label{sec:endpoint}

\subsection{Endpoint series}

Let $g(\tau)=S'(\tau)=\sqrt{\cosh\tau-r}$, which does not vanish for real $\tau$ (there is no real
stationary point). Using
$e^{-2\ii\gamma S}=(\ii/2\gamma g)\,\partial_\tau e^{-2\ii\gamma S}$ and integrating by parts,
\begin{equation}
\int_0^\infty\! f_n\,e^{-2\ii\gamma S}d\tau
=s\,a_n+s\!\int_0^\infty\! f_{n+1}\,e^{-2\ii\gamma S}d\tau ,
\end{equation}
where
\begin{equation}
s=-\frac{\ii}{2\gamma},\qquad f_0=\phi,\quad f_{n+1}=\Big(\frac{f_n}{g}\Big)',\quad a_n=\frac{f_n(0)}{g(0)}.
\end{equation}
The upper limit does not contribute since $f_n/g\to0$ as $\tau\to\infty$. Iterating,
\begin{equation}
\beta_{\rm out}\sim\sum_{n\ge0}s^{n+1}a_n .
\label{eq:series}
\end{equation}

Since $\phi$ is odd and $g$ is even, $f_n$ is odd for even $n$ and even for odd $n$, so
$a_n=0$ for even $n$; in particular $a_0=\phi(0)/g(0)=0$. The first non-zero coefficient follows from
$\phi'(0)=1/[4(1-r)]$ and $g(0)=\sqrt{1-r}$,
\begin{equation}
a_1=\frac{\phi'(0)}{g(0)^2}=\frac1{4(1-r)^2},
\end{equation}
and the next two, obtained by generating $f_n$ symbolically and expanding at $\tau=0$ (SymPy), are
\begin{equation}
a_3=-\frac{r+4}{4(1-r)^4},\qquad a_5=\frac{r^2+26r+55}{4(1-r)^6}.
\end{equation}
Because $s^2=-1/4\gamma^2$, only even powers of $s$ appear and $\beta_{\rm out}$ is real in the phase
convention of Eq.~\eqref{eq:bog} (the Hankel projection of Sect.~\ref{sec:numerics} carries an additional
constant phase):
\begin{align}
\beta_{\rm out}\sim-\frac{1}{16\gamma^2(1-r)^2}\Big[&1+\frac{r+4}{4\gamma^2(1-r)^2}\nonumber\\
&+\frac{r^2+26r+55}{16\gamma^4(1-r)^4}+\dots\Big].
\label{eq:beta}
\end{align}
Hence
\begin{equation}
N_{\rm out}\simeq\frac1{256\,\gamma^4(1-r)^4}\Big[1+\frac{r+4}{2\gamma^2(1-r)^2}+O(\gamma^{-4})\Big].
\label{eq:powerlaw}
\end{equation}
In terms of the initial frequency, using $\Omega_0^2=A(1-r)$,
\begin{equation}
\beta_{\rm out}\simeq-\frac{\kappa^2A}{16\,\Omega_0^4},\qquad
N_{\rm out}\simeq\frac{\kappa^4A^2}{256\,\Omega_0^8}
=\frac{\kappa^4\mu^4C^2}{256(\mu^2C-\Lambda)^4}.
\label{eq:lead}
\end{equation}
Equivalently, $\beta_{\rm out}\simeq-\ddot\Omega(0)/8\Omega_0^3$. The leading contribution is therefore determined entirely by derivatives of $\Omega(t)$ evaluated at the initial endpoint $t=0$. There is no exponential suppression at this order: the result is a pure algebraic endpoint contribution.

\label{sec:turning}

The endpoint series contains all algebraic orders in $1/\gamma$. Contributions beyond this series are exponentially small and are associated with complex zeros of $\Omega^2$, determined by $\cosh\tau_c=r$. For the physical modes $r<-1$, the nearest turning points lie a distance $\pi$ from the real axis in the complex $\tau$ plane. Their contribution is controlled by the imaginary part of the complex action $\gamma S(\tau_c)$, which is of order $\gamma$; it is therefore exponentially small in the large-$\gamma$ limit and lies beyond all algebraic orders in $1/\gamma$. Such contributions are the domain of optimal truncation and the smoothing of Stokes discontinuities \cite{Berry1989,Dabrowski2014,HashibaYamada2021}. Such terms are contained in the exact result of Sect.~\ref{sec:inout}, through the difference between $p_n$ and its adiabatic expansion; we do not separate them here.

\subsection{Dependence on the adiabatic order}
\label{sec:order}

The endpoint series has a direct interpretation in terms of the adiabatic expansion of
Sect.~\ref{sec:adexp}. Each integration by parts in Sect.~\ref{sec:endpoint} brings down one factor of
$1/\gamma$, the adiabatic expansion parameter, and evaluates a combination of time derivatives of
$\Omega$ at the endpoint. The terms $s^{n+1}a_n$ are therefore the mismatch, order by order, between
the zeroth-order adiabatic state imposed at $t=0$ and the exact adiabatic structure of the mode at
that instant. Imposing the state at a higher order removes this mismatch through the corresponding
order. The order-2 adiabatic vacuum uses the explicit truncation \eqref{eq:W2}, which at
$t=0$ (where $\dot\Omega=0$) gives $W_{(2)}=\Omega_0-\ddot\Omega(0)/4\Omega_0^2$ up to higher adiabatic order.
Expanding the initial data $T(0)=(2W_{(2)})^{-1/2}$, $\dot T(0)=-\ii W_{(2)}T(0)$ in the zeroth-order basis gives
$\beta(0)=\ddot\Omega(0)/8\Omega_0^3=+\kappa^2A/16\Omega_0^4$, which cancels the leading term of
\eqref{eq:lead}. The remaining amplitude is $O(\gamma^{-4})$ and $N_{\rm out}$ falls steeply
(numbers in Sect.~\ref{sec:numerics}). More generally, an order-$2k$ vacuum removes the endpoint
terms through $O(\gamma^{-2k})$ in $\beta_{\rm out}$, so that $\beta_{\rm out}=O(\gamma^{-2k-2})$ and
$N_{\rm out}=O(\gamma^{-4k-4})$ for fixed $k$ as $\gamma\to\infty$. For general $k$ this is an asymptotic statement that follows from the exact result \eqref{eq:exact} below, through the property that $W_{(2k)}(0)$ is the truncated adiabatic expansion of $p_n$. We have verified it by direct integration for $k\le1$, and with the closed form for $k\le3$ (Sect.~\ref{sec:numerics}); it has not been tested beyond $k=3$. The leading power law is therefore
the imprint of the order of the adiabatic vacuum chosen at $t=0$. This parallels the situation in
cosmology, where the adiabatic order of the initial state determines the ultraviolet falloff of the
created particle number \cite{ParkerFulling1974,Parker2012}, and where the precision of WKB-based estimates of particle production has been analysed in detail \cite{Winitzki2005}. A finite adiabatic order cancels
successive algebraic endpoint terms but does not remove the whole series. The state for which the
whole series vanishes is the out vacuum itself, whose data at $t=0$ are given in closed form in Sect.~\ref{sec:inout}.

\subsection{Exact result for the particle number}
\label{sec:inout}

The particle number can in fact be obtained exactly, and its large-$\gamma$ expansion reproduces the endpoint series. Since $\cosh(\kappa t+\ii\pi)=-\cosh\kappa t$, the
substitution $x=t+\ii\pi/\kappa$ maps the spatial equation \eqref{eq:X} onto the temporal equation
\eqref{eq:T} with the same $\Lambda$. The coefficients of \eqref{eq:X} are entire, so the eigenfunctions
$X_n$ are entire functions of $x$, and
\begin{equation}
u_n(t)=c_n\,X_n\big(t+\ii\pi/\kappa\big)
\label{eq:cont}
\end{equation}
is an exact solution of the temporal equation of mode $n$ for all real $t$. Its behaviour as $t\to+\infty$
follows from that of $X_n$ as $x\to+\infty$, where $X_n$ is the recessive solution of
$X''\simeq\frac A2e^{\kappa x}X$, proportional to $K_{\ii\sigma_n}(z)$ with
$z=\sqrt2\,\gamma\,e^{\kappa x/2}$, up to corrections that vanish in that limit. The asymptotic form
$K_\nu(z)\simeq\sqrt{\pi/2z}\,e^{-z}$ holds for $|\arg z|<3\pi/2$ \cite[Sec.~10.40]{NISTHandbook}. On the
line ${\rm Im}\,x=\pi/\kappa$ one has $z=\ii\xi$, and $K_\nu(\ii\xi)\propto H^{(2)}_\nu(\xi)$
\cite[Sec.~10.27]{NISTHandbook}, so $u_n$ is purely of positive frequency as $t\to+\infty$: $u_n\simeq{\rm const}\times u_+$, where $u_+$ is the asymptotic Hankel mode of
Eq.~\eqref{eq:out}, with no admixture of the negative-frequency mode $u_-$. It is therefore the exact positive-frequency out mode of the full temporal equation, including the term $\frac A2e^{-\kappa t}$ omitted in Eq.~\eqref{eq:out}.

This gives the particle number in closed form. Along $x=\ii y/\kappa$ the
function $Y_n(y)\propto X_n(\ii y/\kappa)$ can be taken real and satisfies the ordinary Mathieu
equation \cite[Sec.~28.2]{NISTHandbook}
\begin{equation}
Y_n''+\gamma^2\big(\cos y+r_n\big)Y_n=0,
\end{equation}
with $Y_n'(0)=0$ for even $n$ and $Y_n(0)=0$ for odd $n$. At $t=0$ the out mode
\eqref{eq:cont} therefore satisfies $\dot u_n/u_n=-\ii p_n$, with
\begin{equation}
p_n=\kappa\,\frac{Y_n'(\pi)}{Y_n(\pi)} .
\end{equation}
It has the adiabatic form $\dot u=-\ii Wu$ at $t=0$ with $W(0)=p_n$ and $\dot W(0)=0$.

The number $p_n$ is real, positive and finite for every $n$ and $\gamma$. Reality follows from parity: $X_n$ is real on the real axis and has the parity of $n$, so for even $n$ the value $X_n(\ii y/\kappa)$ is real and $X_n'(\ii y/\kappa)$ is purely imaginary, and the opposite holds for odd $n$; in both cases $\dot u_n/u_n$ is purely imaginary at $t=0$. For positivity and finiteness, note that $u_n$ tends to a non-zero multiple $c\,u_+$ of the positive-frequency Hankel mode as $t\to+\infty$. The Wronskian $\mathcal W=u_n\dot u_n^*-u_n^*\dot u_n$ is independent of $t$ and its late-time value is $\ii|c|^2$, since $u_+\dot u_+^*-u_+^*\dot u_+=\ii$. If $Y_n(\pi)$ vanished, then $u_n(0)=0$ and $\mathcal W=0$, a contradiction; hence $Y_n(\pi)\neq0$ and $p_n$ is finite. Normalising $u_n$ so that $\mathcal W=\ii$ and using $\dot u_n(0)=-\ii p_nu_n(0)$ gives $\mathcal W=2\ii p_n|u_n(0)|^2$, hence
\begin{equation}
|u_n(0)|^2=\frac{1}{2p_n},\qquad p_n>0 .
\end{equation}
Section~\ref{sec:minkowski} gives the stronger bound $p_n\ge\sqrt{2A}$. For the initial state
$T(0)=(2W_0)^{-1/2}$, $\dot T(0)=-\ii W_0T(0)$, the coefficient $\beta_{\rm out}=\ii\,(T\dot u_n-\dot Tu_n)$ then gives,
for all $\gamma$,
\begin{equation}
N_{\rm out}=\frac{(W_0-p_n)^2}{4\,W_0\,p_n},
\label{eq:exact}
\end{equation}
with $W_0=\Omega_0$ for the zeroth-order state \eqref{eq:ic} and $W_0=W_{(2k)}(0)$ for the
order-$2k$ state. Since $p_n$ is the exact value at $t=0$ of the function $W$ of Eq.~\eqref{eq:Weq} for
the out mode, the truncations $W_{(2k)}(0)$ are its asymptotic expansion in $1/\gamma$, with
$W_{(2k)}(0)-p_n=O(\gamma^{-2k-1})$. Up to a constant phase, the endpoint series \eqref{eq:beta} is the expansion of
$(W_0-p_n)/2\sqrt{W_0p_n}$. At leading order $p_n\simeq W_{(2)}(0)=\Omega_0-\ddot\Omega(0)/4\Omega_0^2$,
which reproduces Eq.~\eqref{eq:lead}. The half-line particle number thus measures only the mismatch, at
$t=0$, between the prescribed initial state and the out vacuum. A state with $W_0=p_n$ is the out vacuum itself and gives
$N_{\rm out}=0$. Equation \eqref{eq:exact} is checked numerically in
Sect.~\ref{sec:numerics}.

\subsection{The out vacuum is the Minkowski vacuum}
\label{sec:minkowski}

Including the transverse dependence, the out modes \eqref{eq:cont} are
\begin{equation}
\Phi_n(t,x,\bm y)=c_n\,X_n\big(t+\ii\pi/\kappa\big)\,X_n(x)\,e^{\ii\bm k_\perp\cdot\bm y},
\label{eq:outmode}
\end{equation}
with $\bm y=(y,z)$. A solution of the Klein--Gordon equation in the $(\bar t,\bar x)$ plane is of positive frequency with respect to $\bar t$ when it extends to a bounded holomorphic function in the backward tube ${\rm Im}\,\bar u<0$, ${\rm Im}\,\bar v<0$, with $\bar u=\bar t-\bar x$ and $\bar v=\bar t+\bar x$ \cite{StreaterWightman1964}; a positive-frequency plane wave decays there and a negative-frequency one grows.

Continuing $u\to u-\ii a$ and $v\to v-\ii b$ with $0<\kappa a,\kappa b<2\pi$ gives ${\rm Im}\,\bar u=-\ell\cosh(\kappa u/2)\sin(\kappa a/2)<0$, and similarly ${\rm Im}\,\bar v<0$; since $\sinh$ maps the strip $-\pi<{\rm Im}\,w<0$ onto the lower half-plane, these values cover the whole tube. The arguments of both factors $X_n$ then have imaginary parts in $(-\pi/\kappa,\pi/\kappa)$, where $X_n$ is bounded because its asymptotic form $K_{\ii\sigma_n}\big(\sqrt2\,\gamma\,e^{\kappa z/2}\big)$ decays. Each point of the tube has four preimages, related by $u\to-2\pi\ii/\kappa-u$ and $v\to-2\pi\ii/\kappa-v$. The first exchanges the two factors of Eq.~\eqref{eq:outmode}; the second does so up to signs $(-1)^n$ that cancel in the product, so $\Phi_n$ is single valued. The out modes are therefore of positive frequency with respect to Minkowski time, and the out vacuum is the Minkowski vacuum.

On the slice $t=0$, which is the inertial slice $\bar t=0$, $\partial_t=N\,\partial_{\bar t}$ with lapse
\begin{equation}
N=\sqrt{F(0,x)}=\sqrt{2C}\cosh\frac{\kappa x}{2}=\Big(2C+\frac{\kappa^2\bar x^2}{4}\Big)^{1/2}.
\label{eq:lapse}
\end{equation}
Minkowski positive-frequency data obey $\partial_t\phi=-\ii\,\omega_M\phi$, with $\omega_M=N\bar\omega$ and $\bar\omega=(-\partial_{\bar x}^2+\mu^2)^{1/2}$. In the inner product $\langle f,g\rangle=\int dx\,f^*g$, in which the $X_n$ are orthonormal, $\omega_M$ is symmetric because $d\bar x=N\,dx$, and the result above reads
\begin{equation}
\omega_MX_n=p_n\,X_n ,
\label{eq:omegaM}
\end{equation}
so $p_n$ is an eigenvalue of the Minkowski one-particle operator written in the frame; a numerical check is given in the supplementary material. Since $\langle X_n,\omega_MX_n\rangle=\int d\bar x\,X_n\bar\omega X_n\ge\mu\int dx\,NX_n^2$ and $N\ge\sqrt{2C}$, it follows that $p_n\ge\sqrt{2A}$.

The initial state is the ground state of the frame Hamiltonian $H(0)=\int d\bar x\,d^2y\,N\,T_{\bar t\bar t}$, whose one-particle operator $(-\partial_x^2+\mu^2F(0,x))^{1/2}$ has eigenvalues $\Omega_0$ on the same $X_n$. Equation \eqref{eq:exact} is thus the Bogoliubov coefficient between the ground states of two commuting quadratic Hamiltonians, and $N_{\rm out}$ is the number of inertial quanta in the instantaneous vacuum of the accelerated frame. The acceleration enters through the lapse, $\mathfrak a_0=|dN/d\bar x|/N$. A constant $N$ gives the Minkowski Hamiltonian and $N_{\rm out}=0$, while $N\propto|\bar x|$ gives the Rindler boost and the Fulling vacuum, whose Minkowski content is thermal. The lapse \eqref{eq:lapse} interpolates between the two. In terms of the maximal acceleration $\mathfrak a_{\max}=\kappa/(4\sqrt{2C})$ of Sect.~\ref{sec:geometry}, Eq.~\eqref{eq:lead} reads
\begin{equation}
N_{\rm out}\simeq\frac{4\,C^2A^2\,\mathfrak a_{\max}^4}{\Omega_0^8},
\label{eq:Nacc}
\end{equation}
so at fixed $A$, $C$ and $\Omega_0$ the particle number vanishes as the fourth power of the maximal acceleration.

\subsection{Initial instant $t_0\neq0$}
\label{sec:t0}

The construction of Sect.~\ref{sec:inout} applies at any initial instant. The out mode at $t_0$ is $u_n(t_0)\propto X_n(t_0+\ii\pi/\kappa)$, and we write $\dot u_n/u_n=-\ii P_n(t_0)$, with $P_n(0)=p_n$ and ${\rm Re}\,P_n>0$ by the Wronskian argument above. For initial data $T=(2W_0)^{-1/2}$, $\dot T=-\ii VT$ with ${\rm Re}\,V=W_0$, the coefficient $\beta_{\rm out}=\ii(T\dot u_n-\dot Tu_n)$ gives
\begin{equation}
N_{\rm out}=\frac{|V-P_n(t_0)|^2}{4\,W_0\,{\rm Re}\,P_n(t_0)} ,
\label{eq:exact-t0}
\end{equation}
which reduces to Eq.~\eqref{eq:exact} at $t_0=0$. For $t_0\neq0$ the result depends on how the zeroth-order state is defined, because $\dot\Omega(t_0)\neq0$. For the instantaneous ground state of $H(t_0)$, $V=\Omega(t_0)$ as in Eq.~\eqref{eq:ic}. The exact out mode has $P_n=W-\ii\dot W/2W$ up to terms beyond all orders, so the first endpoint term no longer vanishes and
\begin{equation}
\beta_{\rm out}\simeq-\frac{\ii\dot\Omega(t_0)}{4\Omega(t_0)^2},\qquad N_{\rm out}\simeq\frac{\dot\Omega^2}{16\,\Omega^4}=\frac{\epsilon(t_0)^2}{16}=O(\gamma^{-2}) .
\end{equation}
For the zeroth-order WKB state, which keeps the prefactor term, $V=\Omega-\ii\dot\Omega/2\Omega$, the $O(\gamma^{-1})$ term cancels and
\begin{equation}
\beta_{\rm out}\simeq\frac{W_{(2)}-\Omega}{2\Omega}\bigg|_{t_0}=-\frac{\ddot\Omega}{8\Omega^3}+\frac{3\dot\Omega^2}{16\,\Omega^4}\bigg|_{t_0},
\label{eq:beta-t0}
\end{equation}
so that $N_{\rm out}=O(\gamma^{-4})$, with a coefficient that depends on $t_0$. The law $N_{\rm out}\propto\Omega_0^{-8}$ of Eq.~\eqref{eq:lead} is therefore a property of the symmetric instant $t_0=0$, where $\dot\Omega=0$ and the two definitions of the zeroth-order state coincide. At $t_0=0.5/\kappa$ and $1/\kappa$, the exact result \eqref{eq:exact-t0} for the mode $n=0$ approaches both asymptotic forms: at $\gamma=16$ the ratios to the predictions are $1.0027$ and $1.0011$ for $\kappa t_0=0.5$, and $1.0010$ and $1.0003$ for $\kappa t_0=1$, for the instantaneous and WKB states respectively.

\section{Numerical results}
\label{sec:numerics}

\subsection{Methods}

\emph{Direct numerical integration.} We work in units $\kappa=1$, so that $\tau=t$ and the Wronskian normalisation is taken with respect to $\tau$. We integrate the full mode equation $T''+\gamma^2(\cosh\tau-r)T=0$ (prime is $d/d\tau$) from the data \eqref{eq:ic} to $\tau_{\rm end}=11$ (DOP853, relative tolerance $10^{-13}$) and project onto the asymptotic
Hankel out basis \eqref{eq:out} by Wronskians. At $\tau_{\rm end}=11$ the neglected $e^{-\tau}$ branch of $\Omega^2$
contributes only at relative order $e^{-2\tau_{\rm end}}$; the stability of the projection with respect to the
integration endpoint and the numerical tolerances is verified in the supplementary material. The coefficients are
\begin{equation}
\alpha_{\rm out}=\frac{W[T,u_-]}{\ii},\quad \beta_{\rm out}=\ii\,W[T,u_+],
\label{eq:proj}
\end{equation}
with $W[f,g]=fg'-f'g$ and $u_\pm$ the out modes, which solve the temporal equation exactly only after the $\tfrac A2e^{-\kappa t}$ term is dropped. Hankel
functions of real or imaginary order are evaluated with mpmath. No weak-mixing assumption enters the numerical integration,
and the computed coefficients satisfy $|\alpha_{\rm out}|^2-|\beta_{\rm out}|^2=1$ to the working precision of the integration.

Projecting onto the asymptotic Hankel out modes rather than onto the instantaneous
adiabatic basis is essential at this value of $\tau_{\rm end}$. The
zeroth-order adiabatic number still carries a residual oscillation of relative
size $\epsilon/|\beta_{\rm out}|$, which is of order unity at $\tau=11$ for the
modes considered here and decays only as $e^{-\kappa t/2}$; reading
$|\beta|^2$ off the adiabatic basis at $\tau=11$ overestimates $N_{\rm out}$ by
about a third, and the two agree to about $1\%$ only from $\tau\approx13$ onwards. The Hankel basis is
free of this because it solves the asymptotic equation exactly. The physical eigenvalues $r_n(\gamma)$ are obtained numerically by diagonalising the spatial operator $H_x=-\partial_x^2+A\cosh\kappa x$.

\subsection{Results for the physical modes}

\begin{figure}[t]
\includegraphics[width=0.95\textwidth]{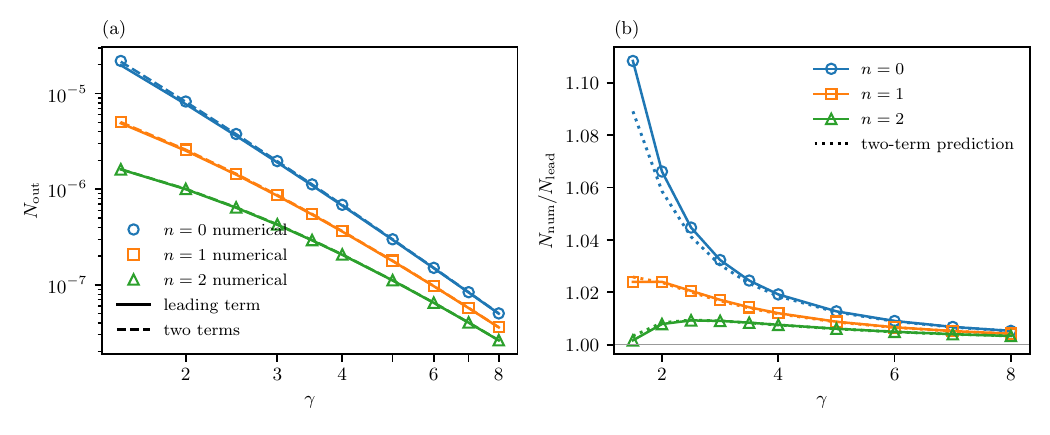}
\caption{Particle number for the adiabatic vacuum imposed at $t=0$, for the physical modes
$r=r_n(\gamma)$, $n=0,1,2$. (a) Numerically integrated $N_{\rm out}$ (points), the leading term of
Eq.~\eqref{eq:powerlaw} (solid) and the two-term series (dashed). (b) Ratio of the numerical to the
leading-order number (points) and the ratio predicted by the two-term series (dotted).}
\label{fig:numerics}
\end{figure}

Figure~\ref{fig:numerics} compares the numerically integrated particle number with the endpoint series
for the three lowest physical modes, $r=r_n(\gamma)$ with $n=0,1,2$, for $1.5\le\gamma\le8$. The ratio
$N_{\rm num}/N_{\rm lead}$ tends to $1$ for every mode, and the approach is the one predicted by the
correction in Eq.~\eqref{eq:powerlaw}. For $n=0$ the ratio is $1.108$ at $\gamma=1.5$, $1.032$ at
$\gamma=3$ and $1.0053$ at $\gamma=8$, against $1.089$, $1.031$ and $1.0052$ from the two-term series.
Because $1-r_n>2$, the leading term is already accurate at moderate $\gamma$, and the agreement improves with $n$: at
$\gamma=3$ the numerical values $N_0=1.969\times10^{-6}$, $N_1=8.700\times10^{-7}$ and
$N_2=4.294\times10^{-7}$ differ from the leading term by $3.2\%$, $1.7\%$ and $0.9\%$, and from the
two-term series by less than $0.2\%$. For large $n$, $N_n$ decreases as $E_n^{-4}$ according to
Eq.~\eqref{eq:lead}.

To test the dependence on adiabatic order, we also evolved states of order $2k$, $k=0,1$, at the representative value $r=-1.5$, chosen within the range $r<-1$ satisfied by physical modes. For $k\ge2$ the predicted $|\beta_{\rm out}|$ falls below the residual of the integration for $\gamma\ge4$, so those orders are not reported. Fitting $|\beta_{\rm out}|$ over $4\le\gamma\le12$ gives exponents $-2.005$ for $k=0$ and $-4.017$ for $k=1$, against the predicted $-2$ and $-4$. The closed form \eqref{eq:exact} is not limited by the integration residual, since it only requires the out-mode data at $t=0$. For a value of $r$ that is not an eigenvalue, these data are obtained from the solution of Eq.~\eqref{eq:X} that decays as $x\to+\infty$, continued to $x=\ii\pi/\kappa$; the argument of Sect.~\ref{sec:inout} uses only this recessive behaviour. With $W_{(2k)}(0)$ computed to order $k=3$ from Eq.~\eqref{eq:Weq}, the fitted exponents of $|\beta_{\rm out}|$ over $4\le\gamma\le12$ at $r=-1.5$ are $-2.005$, $-4.015$, $-6.025$ and $-8.021$ for $k=0,1,2,3$, against the predicted $-2,-4,-6,-8$. This confirms that increasing the adiabatic order cancels successive algebraic endpoint terms, through order $k=3$.

Table~\ref{tab:exact} compares the closed form \eqref{eq:exact} with the direct integration described above. The number $p_n$ is obtained by integrating the Mathieu equation for $Y_n$ over $0\le y\le\pi$ with the eigenvalue $r_n(\gamma)$. For the zeroth-order state the two agree to within one unit in the seventh significant digit. For the second-order state they agree to better than $3\times10^{-4}$, the difference being the error of the direct integration at these small values.

Over $0.25\le\gamma\le12$ and $0\le n\le9$, $p_n$ is positive and finite, with smallest value $p_0=0.531$ at $\gamma=0.25$, consistent with the bound $p_n\ge\sqrt{2A}$ of Sect.~\ref{sec:minkowski}.

\emph{Total number.} For $\mu^2=m^2+k_\perp^2$, the number of created quanta per unit transverse area is
\begin{equation}
n_\perp=\sum_n\int\frac{d^2k_\perp}{(2\pi)^2}\,N_n=\frac{1}{4\pi C}\int_{m^2C}^{\infty}dA\,\sum_nN_n(A).
\end{equation}
At large $A$ the sum over $n$ is dominated by levels with $E_n-A$ of order $A$, beyond the harmonic region at the bottom of the potential. Replacing the sum by an integral over $E$ with the semiclassical density of levels of $-\partial_x^2+A\cosh\kappa x$, and using Eq.~\eqref{eq:lead} with $\Omega_0^2=A+E_n$, gives $\sum_nN_n\simeq25\pi\kappa^3/(2^{19}\sqrt2\,A^{3/2})\approx1.06\times10^{-4}\,\kappa^3A^{-3/2}$. The integrand therefore falls as $A^{-3/2}$, and $n_\perp$ is finite for $m>0$. Evaluating the sum with Eq.~\eqref{eq:exact} for $n\le40$ and $A\le400$, and with the leading-order law for the remaining modes, gives $n_\perp\simeq1.63\times10^{-5}$ for $\kappa=C=m=1$, accurate to about $2\%$. The numerically fitted large-$A$ exponent is $-1.50$, and for $100\le A\le3000$ the sum agrees with the asymptotic form to $0.1\%$.

\emph{Total energy.} The energy carried by these quanta behaves differently. A quantum in the mode $(n,\bm k_\perp)$ has Minkowski energy $p_n$, Eq.~\eqref{eq:omegaM}, and $p_n\simeq\Omega_0$ at leading order, so the same calculation gives $\sum_np_nN_n\simeq\kappa^3/(1800\pi A)$. The energy per unit transverse area, $\varepsilon_\perp=(4\pi C)^{-1}\int dA\,\sum_np_nN_n$, therefore diverges logarithmically at large transverse momentum: with a cutoff $k_{\max}\gg m$, $\varepsilon_\perp\simeq\kappa^3\ln(k_{\max}/m)/(3600\pi^2C)$. Numerically, the fitted large-$A$ exponent of $\sum_np_nN_n$ is $-1.00$, and $A\sum_np_nN_n$ approaches the predicted constant to $0.1\%$. For the second-order state $N_{\rm out}=O(\gamma^{-8})$ and $\varepsilon_\perp$ is finite. This is the integrated counterpart of the statement in Sect.~\ref{sec:discussion} that a finite local $\langle T_{\mu\nu}\rangle$ requires a state of higher adiabatic order.

\begin{table}[t]
\caption{Closed form \eqref{eq:exact} against direct integration for the zeroth-order ($k=0$) and second-order ($k=1$) adiabatic states imposed at $t=0$ ($\kappa=1$). Numbers $N$ are in units of $10^{-7}$ for $k=0$ and $10^{-12}$ for $k=1$.}
\label{tab:exact}
\begin{tabular}{@{}cccccccc@{}}
\toprule
$\gamma$ & $n$ & $r_n$ & $p_n$ & $N^{(k=0)}_{\rm exact}$ & $N^{(k=0)}_{\rm num}$ & $N^{(k=1)}_{\rm exact}$ & $N^{(k=1)}_{\rm num}$\\
\midrule
1.5 & 0 & $-1.497778$ & 2.348533 & 219.7155 & 219.7155 & 37979.56 & 37979.54\\
3 & 0 & $-1.242458$ & 4.479863 & 19.68773 & 19.68773 & 335.4907 & 335.4898\\
3 & 1 & $-1.740214$ & 4.956822 & 8.699857 & 8.699857 & 38.07892 & 38.07924\\
3 & 2 & $-2.262804$ & 5.411873 & 4.293763 & 4.293762 & 4.423882 & 4.423770\\
8 & 0 & $-1.089354$ & 11.55849 & 0.5030719 & 0.5030717 & 0.2376100 & 0.2375477\\
\botrule
\end{tabular}
\end{table}

\section{Discussion}
\label{sec:discussion}

\emph{Algebraic rather than Planckian spectrum.} For the state prepared at $t=0$, Eq.~\eqref{eq:lead} gives $N_{\rm out}\propto\Omega_0^{-8}$ at leading order. A Bose--Einstein occupation decreases exponentially with the frequency at large $\omega/T$, whereas the present result is a power law. The statement is limited in scope. It concerns one state, the adiabatic vacuum imposed at $t=0$, and one set of mode labels, the out modes $(n,\bm k_\perp)$ of the frame. It does not exclude a thermal-looking response of a detector over a finite interval, such as the local temperature found in Ref.~\cite{Sobrero2025}, nor does it say how the same quanta would be distributed over a different set of Minkowski modes. In any case, a temperature in the strict sense requires a mixed state. In the Rindler case this arises from tracing over the modes beyond the horizon, whereas here, as for the asymptotically inertial observers of Ref.~\cite{MunozDoria2026}, there is no horizon and nothing is traced out. The state prepared at $t=0$ is a pure Gaussian state, related to the Minkowski vacuum by a Bogoliubov transformation diagonal in the mode labels. It is not a calculation of the response of an Unruh--DeWitt detector.

\emph{Why no Rindler horizon appears.} The coordinate transformation \eqref{eq:embedding} makes the geometry transparent. For the future evolution considered here, each fixed-$x$ observer has a proper acceleration that decreases to zero and a velocity that approaches the finite limit \eqref{eq:vel}. No worldline approaches the light cone, and there is no Rindler horizon. The temporal mode frequency remains positive on the real half-line for every physical mode, so the problem is naturally a finite-time, time-dependent-frequency problem rather than a horizon-crossing one.

\emph{What $N_{\rm out}$ measures.} Because the out vacuum is the Minkowski vacuum (Sect.~\ref{sec:minkowski}), $N_{\rm out}$ is the number of inertial quanta already contained in the state prepared at $t=0$; the late-time evolution, which is adiabatic, adds essentially nothing. The acceleration enters through the choice of that state, via the position-dependent lapse \eqref{eq:lapse}, and Eq.~\eqref{eq:Nacc} makes the dependence explicit.

\emph{Total number and stress tensor.} The total number per unit transverse area is finite, but for the zeroth-order state the total energy per unit transverse area diverges logarithmically (Sect.~\ref{sec:numerics}). The renormalised stress tensor in the late inertial region is $\langle T_{\mu\nu}\rangle_{\rm ren}=\langle{:}T_{\mu\nu}{:}\rangle$, normal ordered with respect to the Minkowski, equivalently out, modes. For the zeroth-order state this expectation value is not finite. Since $|\beta_{\rm out}|$ decreases only as $k_\perp^{-2}$ at large transverse momentum, the local energy density diverges, and its integral, the energy per unit transverse area, diverges logarithmically (Sect.~\ref{sec:numerics}). This agrees with the adiabatic counting of Parker and Fulling, in which a finite $\langle T_{\mu\nu}\rangle$ requires an initial state of adiabatic order four or higher \cite{ParkerFulling1974,LudersRoberts1990}. For such states $|\beta_{\rm out}|=O(\gamma^{-6})$, the mode sums converge and $\langle T_{\mu\nu}\rangle_{\rm ren}$ is finite; its explicit evaluation is not attempted here.

\emph{Relation to switching effects in detector models.} The structure found here has a counterpart in the theory of particle detectors, which we state as an analogy rather than as a derived equivalence. The response of an Unruh--DeWitt detector that is switched on at a finite time contains transient contributions determined by the switching, which can dominate the response unless the switching is smooth \cite{SriramkumarPadmanabhan1996,LoukoSatz2006,Satz2007,Stargen2026}. In the present field-mode calculation the role of the switching is played by the instant $t=0$ at which the state is imposed: the leading amplitude $\beta_{\rm out}\simeq-\ddot\Omega(0)/8\Omega_0^3$ depends only on local data at that instant. Raising the adiabatic order of the initial state is the analogue of smoothing the switching function: it suppresses the transient order by order. The closed form shows where this process ends: the adiabatic truncations approach the out vacuum, the state with $W_0=p_n$, for which $N_{\rm out}=0$.

A direct detector calculation would make this analogy quantitative. For a detector at fixed $x$ in the Minkowski vacuum, the response along these worldlines has been computed in Ref.~\cite{Sobrero2025}: it is thermal-like over a finite interval and becomes inertial at late times. By the result of Sect.~\ref{sec:minkowski}, that vacuum is the out vacuum of the present problem. The response in the state prepared at $t=0$ is the response in the Minkowski vacuum plus terms determined by the Bogoliubov coefficients of Sect.~\ref{sec:inout}: the occupation $N_{\rm out}=|\beta_{\rm out}|^2$ and the interference terms in $\alpha_{\rm out}\beta_{\rm out}^*$, which arise because the prepared state is squeezed relative to the Minkowski vacuum. Evaluating these terms requires the mode sum over $n$ and $\bm k_\perp$ along the worldline, which we have not carried out here.

\emph{Comparison with asymptotically uniformly accelerated frames.} In the frames studied in Refs.~\cite{PercocoVillalba1992,VillalbaMateu2000} the observers approach uniform acceleration at late times, which is the regime in which a Rindler horizon and the Unruh effect arise. The present frame realises the opposite asymptotics: the proper acceleration decays as the inverse cube of proper time, the total rapidity acquired by each observer is finite, and no horizon forms. Consistently, the particle number obtained here is algebraic rather than thermal.

\section{Conclusions}
\label{sec:conclusions}

We have studied the forward evolution of a massive scalar field from $t=0$ to $t\to+\infty$ in a time-dependent accelerated coordinate system in flat spacetime. At $t=0$ the fixed-coordinate observers have their maximum proper acceleration; for $t>0$ the acceleration decreases monotonically and tends to zero, so the observers become asymptotically inertial.

The Klein--Gordon equation separates into a modified Mathieu equation in time and a confining Schr\"odinger problem in space. The physical separation constants are discrete and satisfy $\Lambda_n<-A$. Imposing an adiabatic vacuum at the finite initial time $t=0$ gives an endpoint expansion for the future particle number. For the zeroth-order state the leading term is
\begin{equation}
N_{\rm out}\simeq\frac{\kappa^4A^2}{256\Omega_0^8},
\end{equation}
in agreement with direct numerical integration of the physical modes. Increasing the adiabatic order removes successive algebraic endpoint terms and produces a correspondingly stronger suppression in the adiabatic large-$\gamma$ limit. The leading particle number is thus a transient set by the instant at which the state is prescribed, the field-mode analogue of the switching transients familiar from finite-time particle detectors.

Continuing the spatial eigenfunctions to ${\rm Im}\,x=\pi/\kappa$ gives the positive-frequency out modes exactly, and hence the exact result $N_{\rm out}=(W_0-p_n)^2/4W_0p_n$, with $p_n$ obtained from an ordinary Mathieu equation on $0\le y\le\pi$. The endpoint series is its expansion in $1/\gamma$. The out modes are of positive frequency with respect to Minkowski time, so the out vacuum is the Minkowski vacuum and $p_n$ is an eigenvalue of the Minkowski one-particle operator on the slice $t=0$. The particle number is therefore the number of inertial quanta in the instantaneous vacuum of the accelerated frame; it scales as the fourth power of the maximal acceleration and is finite when summed over all modes, although for the zeroth-order state the energy carried by these quanta diverges logarithmically at large transverse momentum. The $\Omega_0^{-8}$ law is specific to the symmetric instant $t=0$; at other initial instants the leading power depends on how the zeroth-order state is defined.

For this state and these mode labels the occupation numbers decrease algebraically rather than exponentially. The essential point is that this is a finite-time initial-value problem: the state is prescribed at the instant of maximum acceleration and evolved toward a future in which the acceleration vanishes. No assumption about the state at $t\to-\infty$ is required.

\backmatter

\bmhead{Supplementary information}
The scripts used for the numerical results of Sects.~\ref{sec:endpoint} and \ref{sec:numerics} are provided as supplementary material, with the data underlying Fig.~\ref{fig:numerics} and the values quoted in the text printed as tables. They compute the spectrum of the spatial operator, integrate the temporal mode equation, project onto the asymptotic Hankel basis, evaluate the closed forms \eqref{eq:exact} and \eqref{eq:exact-t0}, reproduce the adiabatic-order scaling of Sect.~\ref{sec:order}, check Eq.~\eqref{eq:omegaM}, and compute the total number $n_\perp$ and the transverse energy sum.

\section*{Declarations}

\begin{itemize}
\item \textbf{Funding.} This work received no specific grant from any funding agency.
\item \textbf{Conflict of interest.} The author declares no conflict of interest.
\item \textbf{Data availability.} The data underlying Fig.~\ref{fig:numerics} and the numerical values quoted in the text are provided as supplementary material.
\item \textbf{Code availability.} The code used to generate the numerical results is provided as supplementary material.
\end{itemize}

\bibliography{references}

@article{Fulling1973,
  author  = {Fulling, S. A.},
  title   = {Nonuniqueness of canonical field quantization in {Riemannian} space-time},
  journal = {Phys. Rev. D},
  volume  = {7},
  pages   = {2850--2862},
  year    = {1973},
  doi     = {10.1103/PhysRevD.7.2850}
}

@article{Davies1975,
  author  = {Davies, P. C. W.},
  title   = {Scalar production in {Schwarzschild} and {Rindler} metrics},
  journal = {J. Phys. A: Math. Gen.},
  volume  = {8},
  pages   = {609--616},
  year    = {1975},
  doi     = {10.1088/0305-4470/8/4/022}
}

@article{Unruh1976,
  author  = {Unruh, W. G.},
  title   = {Notes on black-hole evaporation},
  journal = {Phys. Rev. D},
  volume  = {14},
  pages   = {870--892},
  year    = {1976},
  doi     = {10.1103/PhysRevD.14.870}
}

@article{Rindler1966,
  author  = {Rindler, W.},
  title   = {Kruskal space and the uniformly accelerated frame},
  journal = {Am. J. Phys.},
  volume  = {34},
  pages   = {1174--1178},
  year    = {1966},
  doi     = {10.1119/1.1972547}
}

@article{UnruhWald1984,
  author  = {Unruh, W. G. and Wald, R. M.},
  title   = {What happens when an accelerating observer detects a {Rindler} particle},
  journal = {Phys. Rev. D},
  volume  = {29},
  pages   = {1047},
  year    = {1984},
  doi     = {10.1103/PhysRevD.29.1047}
}

@article{Takagi1986,
  author  = {Takagi, S.},
  title   = {Vacuum noise and stress induced by uniform acceleration: {Hawking-Unruh} effect in {Rindler} manifold of arbitrary dimension},
  journal = {Prog. Theor. Phys. Suppl.},
  volume  = {88},
  pages   = {1--142},
  year    = {1986},
  doi     = {10.1143/PTPS.88.1}
}

@article{Crispino2008,
  author  = {Crispino, L. C. B. and Higuchi, A. and Matsas, G. E. A.},
  title   = {The {Unruh} effect and its applications},
  journal = {Rev. Mod. Phys.},
  volume  = {80},
  pages   = {787--838},
  year    = {2008},
  doi     = {10.1103/RevModPhys.80.787}
}

@article{SriramkumarPadmanabhan1996,
  author  = {Sriramkumar, L. and Padmanabhan, T.},
  title   = {Finite-time response of inertial and uniformly accelerated {Unruh--DeWitt} detectors},
  journal = {Class. Quantum Grav.},
  volume  = {13},
  pages   = {2061--2079},
  year    = {1996},
  doi     = {10.1088/0264-9381/13/8/005}
}

@article{LoukoSatz2006,
  author  = {Louko, J. and Satz, A.},
  title   = {How often does the {Unruh--DeWitt} detector click? {Regularization} by a spatial profile},
  journal = {Class. Quantum Grav.},
  volume  = {23},
  pages   = {6321--6343},
  year    = {2006},
  doi     = {10.1088/0264-9381/23/22/015}
}

@article{Satz2007,
  author  = {Satz, A.},
  title   = {Then again, how often does the {Unruh--DeWitt} detector click if we switch it carefully?},
  journal = {Class. Quantum Grav.},
  volume  = {24},
  pages   = {1719--1731},
  year    = {2007},
  doi     = {10.1088/0264-9381/24/7/003}
}

@article{ObadiaMilgrom2007,
  author  = {Obadia, N. and Milgrom, M.},
  title   = {Unruh effect for general trajectories},
  journal = {Phys. Rev. D},
  volume  = {75},
  pages   = {065006},
  year    = {2007},
  doi     = {10.1103/PhysRevD.75.065006}
}

@article{KothawalaPadmanabhan2010,
  author  = {Kothawala, D. and Padmanabhan, T.},
  title   = {Response of {Unruh--DeWitt} detector with time-dependent acceleration},
  journal = {Phys. Lett. B},
  volume  = {690},
  pages   = {201--206},
  year    = {2010},
  doi     = {10.1016/j.physletb.2010.05.026}
}

@article{BarbadoVisser2012,
  author  = {Barbado, L. C. and Visser, M.},
  title   = {{Unruh--DeWitt} detector event rate for trajectories with time-dependent acceleration},
  journal = {Phys. Rev. D},
  volume  = {86},
  pages   = {084011},
  year    = {2012},
  doi     = {10.1103/PhysRevD.86.084011}
}

@article{JuarezAubryLouko2014,
  author  = {Ju\'arez-Aubry, B. A. and Louko, J.},
  title   = {Onset and decay of the 1+1 {Hawking--Unruh} effect: what the derivative-coupling detector saw},
  journal = {Class. Quantum Grav.},
  volume  = {31},
  pages   = {245007},
  year    = {2014},
  doi     = {10.1088/0264-9381/31/24/245007}
}

@article{Fewster2016,
  author  = {Fewster, C. J. and Ju\'arez-Aubry, B. A. and Louko, J.},
  title   = {Waiting for {Unruh}},
  journal = {Class. Quantum Grav.},
  volume  = {33},
  pages   = {165003},
  year    = {2016},
  doi     = {10.1088/0264-9381/33/16/165003}
}

@article{Dickinson2025,
  author  = {Dickinson, R. and Forshaw, J. and Jenkinson, R. and Millington, P.},
  title   = {A new study of the {Unruh} effect},
  journal = {Class. Quantum Grav.},
  volume  = {42},
  pages   = {025014},
  year    = {2025},
  doi     = {10.1088/1361-6382/ad9c12}
}

@article{Parry2025,
  author  = {Parry, L. J. A. and Vidal-Cruzprieto, D. and Fewster, C. J. and Louko, J.},
  title   = {Waiting around for {Unruh}},
  journal = {Class. Quantum Grav.},
  volume  = {42},
  pages   = {245012},
  year    = {2025},
  doi     = {10.1088/1361-6382/ae2377}
}

@article{FullingDavies1976,
  author  = {Fulling, S. A. and Davies, P. C. W.},
  title   = {Radiation from a moving mirror in two dimensional space-time: conformal anomaly},
  journal = {Proc. R. Soc. Lond. A},
  volume  = {348},
  pages   = {393--414},
  year    = {1976},
  doi     = {10.1098/rspa.1976.0045}
}

@article{DaviesFulling1977,
  author  = {Davies, P. C. W. and Fulling, S. A.},
  title   = {Radiation from moving mirrors and from black holes},
  journal = {Proc. R. Soc. Lond. A},
  volume  = {356},
  pages   = {237--257},
  year    = {1977},
  doi     = {10.1098/rspa.1977.0130}
}

@article{DuarteBaptista2026,
  author  = {{Duarte-Baptista}, P. and {Va\~{n}\'{o}-Vi\~{n}uales}, A. and {del R\'{\i}o}, A.},
  title   = {A numerical approach to particle creation in accelerating toy models},
  journal = {Class. Quantum Grav.},
  volume  = {43},
  pages   = {045009},
  year    = {2026},
  doi     = {10.1088/1361-6382/ae4201}
}

@article{Costa1989,
  author  = {Costa, I.},
  title   = {Separable coordinates and particle creation. {II}: {Two} new vacua related to accelerating observers},
  journal = {J. Math. Phys.},
  volume  = {30},
  pages   = {888--891},
  year    = {1989},
  doi     = {10.1063/1.528356}
}

@article{PercocoVillalba1992,
  author  = {Percoco, U. and Villalba, V. M.},
  title   = {Particle creation in an asymptotically uniformly accelerated frame},
  journal = {Class. Quantum Grav.},
  volume  = {9},
  pages   = {307--316},
  year    = {1992},
  doi     = {10.1088/0264-9381/9/1/027}
}

@article{VillalbaMateu2000,
  author  = {Villalba, V. M. and Mateu, J.},
  title   = {Vacuum effects in an asymptotically uniformly accelerated frame with a constant magnetic field},
  journal = {Phys. Rev. D},
  volume  = {61},
  pages   = {025007},
  year    = {2000},
  doi     = {10.1103/PhysRevD.61.025007}
}

@article{Parker1969,
  author  = {Parker, L.},
  title   = {Quantized fields and particle creation in expanding universes. {I}},
  journal = {Phys. Rev.},
  volume  = {183},
  pages   = {1057--1068},
  year    = {1969},
  doi     = {10.1103/PhysRev.183.1057}
}

@article{ParkerFulling1974,
  author  = {Parker, L. and Fulling, S. A.},
  title   = {Adiabatic regularization of the energy-momentum tensor of a quantized field in homogeneous spaces},
  journal = {Phys. Rev. D},
  volume  = {9},
  pages   = {341--354},
  year    = {1974},
  doi     = {10.1103/PhysRevD.9.341}
}

@book{BirrellDavies1982,
  author    = {Birrell, N. D. and Davies, P. C. W.},
  title     = {Quantum Fields in Curved Space},
  publisher = {Cambridge University Press},
  address   = {Cambridge},
  year      = {1982},
  doi       = {10.1017/CBO9780511622632}
}

@book{ParkerToms2009,
  author    = {Parker, L. and Toms, D.},
  title     = {Quantum Field Theory in Curved Spacetime: Quantized Fields and Gravity},
  publisher = {Cambridge University Press},
  address   = {Cambridge},
  year      = {2009},
  doi       = {10.1017/CBO9780511813924}
}

@article{LudersRoberts1990,
  author  = {L\"uders, C. and Roberts, J. E.},
  title   = {Local quasiequivalence and adiabatic vacuum states},
  journal = {Commun. Math. Phys.},
  volume  = {134},
  pages   = {29--63},
  year    = {1990},
  doi     = {10.1007/BF02102088}
}

@article{Fulling1979,
  author  = {Fulling, S. A.},
  title   = {Remarks on positive frequency and {Hamiltonians} in expanding universes},
  journal = {Gen. Relativ. Gravit.},
  volume  = {10},
  pages   = {807--824},
  year    = {1979},
  doi     = {10.1007/BF00756661}
}

@article{Parker2012,
  author  = {Parker, L.},
  title   = {Particle creation and particle number in an expanding universe},
  journal = {J. Phys. A: Math. Theor.},
  volume  = {45},
  pages   = {374023},
  year    = {2012},
  doi     = {10.1088/1751-8113/45/37/374023}
}

@article{Winitzki2005,
  author  = {Winitzki, S.},
  title   = {Cosmological particle production and the precision of the {WKB} approximation},
  journal = {Phys. Rev. D},
  volume  = {72},
  pages   = {104011},
  year    = {2005},
  doi     = {10.1103/PhysRevD.72.104011}
}

@article{Dabrowski2014,
  author  = {Dabrowski, R. and Dunne, G. V.},
  title   = {Superadiabatic particle number in {Schwinger} and de {Sitter} particle production},
  journal = {Phys. Rev. D},
  volume  = {90},
  pages   = {025021},
  year    = {2014},
  doi     = {10.1103/PhysRevD.90.025021}
}

@article{Dabrowski2016,
  author  = {Dabrowski, R. and Dunne, G. V.},
  title   = {Time dependence of adiabatic particle number},
  journal = {Phys. Rev. D},
  volume  = {94},
  pages   = {065005},
  year    = {2016},
  doi     = {10.1103/PhysRevD.94.065005}
}

@article{Ilderton2022,
  author  = {Ilderton, A.},
  title   = {Physics of adiabatic particle number in the {Schwinger} effect},
  journal = {Phys. Rev. D},
  volume  = {105},
  pages   = {016021},
  year    = {2022},
  doi     = {10.1103/PhysRevD.105.016021}
}

@article{Berry1989,
  author  = {Berry, M. V.},
  title   = {Uniform asymptotic smoothing of {Stokes's} discontinuities},
  journal = {Proc. R. Soc. Lond. A},
  volume  = {422},
  pages   = {7--21},
  year    = {1989},
  doi     = {10.1098/rspa.1989.0018}
}

@article{KLS1997,
  author  = {Kofman, L. and Linde, A. and Starobinsky, A. A.},
  title   = {Towards the theory of reheating after inflation},
  journal = {Phys. Rev. D},
  volume  = {56},
  pages   = {3258--3295},
  year    = {1997},
  doi     = {10.1103/PhysRevD.56.3258}
}

@book{NISTHandbook,
  editor    = {Olver, F. W. J. and Lozier, D. W. and Boisvert, R. F. and Clark, C. W.},
  title     = {{NIST} Handbook of Mathematical Functions},
  publisher = {Cambridge University Press},
  address   = {Cambridge},
  year      = {2010}
}

@article{Ostapchuk2012,
  author  = {Ostapchuk, D. C. M. and Lin, S.-Y. and Mann, R. B. and Hu, B. L.},
  title   = {Entanglement dynamics between inertial and non-uniformly accelerated detectors},
  journal = {J. High Energy Phys.},
  volume  = {2012},
  number  = {07},
  pages   = {072},
  year    = {2012},
  doi     = {10.1007/JHEP07(2012)072}
}

@article{Sobrero2025,
  author  = {Sobrero, F. and Soares, M. S. and Guerreiro, T. and Svaiter, N. F.},
  title   = {Time-dependent accelerated {Unruh--DeWitt} detector without event horizon},
  journal = {Class. Quantum Grav.},
  volume  = {42},
  pages   = {135013},
  year    = {2025},
  doi     = {10.1088/1361-6382/add707}
}

@article{Stargen2026,
  author  = {Stargen, D. J.},
  title   = {Finite-time {Unruh} effect: Waiting for the transient effects to fade off},
  journal = {Phys. Rev. D},
  volume  = {113},
  pages   = {065007},
  year    = {2026},
  doi     = {10.1103/hqnl-m16b}
}

@article{GoodAndersonEvans2013,
  author  = {Good, M. R. R. and Anderson, P. R. and Evans, C. R.},
  title   = {Time dependence of particle creation from accelerating mirrors},
  journal = {Phys. Rev. D},
  volume  = {88},
  pages   = {025023},
  year    = {2013},
  doi     = {10.1103/PhysRevD.88.025023}
}

@article{Nicolaevici2015,
  author  = {Nicolaevici, N.},
  title   = {Unruh effect without {Rindler} horizon},
  journal = {Class. Quantum Grav.},
  volume  = {32},
  pages   = {045013},
  year    = {2015},
  doi     = {10.1088/0264-9381/32/4/045013}
}

@article{Dubey2026,
  author  = {Dubey, N. K. and Uppalapati, J. and Kolekar, S.},
  title   = {Quantum field theory in successive {Rindler} spacetimes},
  journal = {Phys. Rev. D},
  volume  = {114},
  pages   = {025014},
  year    = {2026},
  doi     = {10.1103/jygq-w4kb}
}

@article{ShahKolekar2026,
  author  = {Shah, H. and Kolekar, S.},
  title   = {Kinematics of acceleration-induced excitations in confined quantum fields},
  journal = {Phys. Rev. D},
  volume  = {113},
  pages   = {045006},
  year    = {2026},
  doi     = {10.1103/2sss-by9v}
}

@article{Agullo2015,
  author  = {Agullo, I. and Nelson, W. and Ashtekar, A.},
  title   = {Preferred instantaneous vacuum for linear scalar fields in cosmological space-times},
  journal = {Phys. Rev. D},
  volume  = {91},
  pages   = {064051},
  year    = {2015},
  doi     = {10.1103/PhysRevD.91.064051}
}

@article{Bianchi2026,
  author  = {Bianchi, E. and Ghelem, Y. and Hackl, L.},
  title   = {Adiabatic vacua from linear complex structures},
  journal = {Phys. Rev. D},
  volume  = {113},
  pages   = {085003},
  year    = {2026},
  doi     = {10.1103/5v25-gczj}
}

@article{HashibaYamada2021,
  author  = {Hashiba, S. and Yamada, Y.},
  title   = {Stokes phenomenon and gravitational particle production---How to evaluate it in practice},
  journal = {J. Cosmol. Astropart. Phys.},
  volume  = {2021},
  number  = {05},
  pages   = {022},
  year    = {2021},
  doi     = {10.1088/1475-7516/2021/05/022}
}

@book{StreaterWightman1964,
  author    = {Streater, R. F. and Wightman, A. S.},
  title     = {{PCT}, Spin and Statistics, and All That},
  publisher = {Benjamin},
  address   = {New York},
  year      = {1964}
}

@article{MunozDoria2026,
  author  = {Mu{\~n}oz, G. and Doria, A.},
  title   = {Does acceleration imply a thermal vacuum?},
  journal = {Ann. Phys.},
  volume  = {490},
  pages   = {170501},
  year    = {2026},
  doi     = {10.1016/j.aop.2026.170501}
}

\end{document}